\documentclass[conference]{IEEEtran}
\IEEEoverridecommandlockouts
\usepackage{cite}
\usepackage{amsmath,amssymb,amsfonts}
\usepackage{BOONDOX-uprscr}
\usepackage{algorithmic}
\usepackage{graphicx}
\usepackage{xcolor}
\usepackage{url}
\usepackage{tabularx,booktabs}
\usepackage{multirow}
\usepackage{threeparttable}
\usepackage{tikz}
\usepackage{textcomp}
\usepackage{hyperref}
\usepackage{lipsum}

\newcommand\copyrighttext{%
    \footnotesize \textbf{This paper has been accepted in the IEEE TRUSTCOM 2021 (https://doi.org/10.1109/TrustCom53373.2021.00152)}\\\\\textcopyright 2021 IEEE. Personal use of this material is permitted. Permission from IEEE must be obtained for all other uses, in any current or future media, including reprinting/republishing this material for advertising or promotional purposes, creating new collective works, for resale or redistribution to servers or lists, or reuse of any copyrighted component of this work in other works.}

\newcommand\copyrightnotice{%
\begin{tikzpicture}[remember picture,overlay]
\node[anchor=south,yshift=10pt] at (current page.south) {\fbox{\parbox{\dimexpr\textwidth-\fboxsep-\fboxrule\relax}{\copyrighttext}}};
\end{tikzpicture}%
}

\def\BibTeX{{\rm B\kern-.05em{\sc i\kern-.025em b}\kern-.08em
    T\kern-.1667em\lower.7ex\hbox{E}\kern-.125emX}}

\begin{document}

\title{\huge{Hop-by-hop Accounting and Rewards for Packet dIspAtching}}

\author{
    \IEEEauthorblockN{Caciano Machado\IEEEauthorrefmark{1}, Renan R. S. dos Santos\IEEEauthorrefmark{2} and Carla Merkle Westphall\IEEEauthorrefmark{3}}
    \IEEEauthorblockA{
        Network and Management Laboratory -- LRG\\
        Graduate Program in Computer Science -- PPGCC\\
        Federal University of Santa Catarina -- UFSC\\
        PO Box 476, 88040-900 -- Florianópolis, Brazil}
    {\small caciano.machado at ufrgs.br\IEEEauthorrefmark{1}, renan.rocha at grad.ufsc.br\IEEEauthorrefmark{2}, carla.merkle.westphall at ufsc.br\IEEEauthorrefmark{3}}
}

\maketitle
\copyrightnotice

\begin{abstract}
Community networks are prone to free-riders, i.e., participants who take
    advantage of cooperation from others' routers but do not contribute
    reciprocally. In this paper, we present HARPIA, a system for credit-based
    incentive mechanisms for data forwarding in community networks aimed to
    prevent selfish behavior. HARPIA does not require a trusted third-party
    or tamper-resistant security modules as in other incentive mechanisms. Instead, it uses
    a distributed accounting scheme (DPIFA) to estimate the balance of data
    forwarding contribution and consumption of each network router and settle
    correspondent cryptocurrency debts on an Ethereum smart contract. On-chain
    settlement transactions are performed every HARPIA cycle (e.g., daily,
    weekly, monthly) and must be validated by at least m-of-n network
    routers using a multi-signature scheme (MuSig). We also realized a
    performance evaluation, security threat assessment, and cryptocurrency costs
    estimation. Results show that our proposal is suitable for community
    networks with up to 64 infrastructure routers under specific m-of-n
    MuSig thresholds.
\end{abstract}


\section{Introduction}
Despite the huge Internet expansion since the ARPANET, there are many regions
and populations in the world that still lack connectivity. Today, the digital
divide is another factor that deepens socioeconomic inequality. In community
networks, to fill the gap left by insufficient market-based and
state-sponsored Internet access solutions, community members deploy and
operate the network infrastructure~\cite{surveycommunity2018}. Still, community networks are prone to the
free-rider problem~\cite{stimulatingcommunity2006}, i.e., routers from selfish participants
who take advantage of the cooperation from others' routers but do not
contribute reciprocally. Free-riders arbitrarily discard or delay packets from
others' routers to save network, computing, and energy resources. This behavior
tends to undermine network reliability and inhibit its expansion. Cooperation
enforcement (or incentive) mechanisms~\cite{surveycoopmanet2006} have been proposed to mitigate potential
free-riders in computer networks. Those mechanisms adopt credit-based,
reputation-based, or algorithmic game theory to encourage cooperative behavior
among routers.

Credit-based mechanisms model the data-forwarding task as a
service valuated and charged using a virtual currency to
regulate the dealings among various routers in multi-hop networks.
Two approaches have been adopted for secure credit-based mechanisms:
tamper-resistant security modules (TRSM) attached to the network interfaces that secure
credit accounting and virtual banks that depend
on a trusted third-party (TTP) service for centralized accounting.
Recently, blockchains started to be adopted in credit-based incentive
mechanisms to manage virtual coins in a trustless based approach, i.e.,
eliminating the need for centralized elements or TRSM. Though, existing
proposals present significant limitations. The main limitations found are the
need for a TTP or frequent expensive on-chain transactions.

\begin{figure}[!htb]
\centering
    \includegraphics[width=\columnwidth]{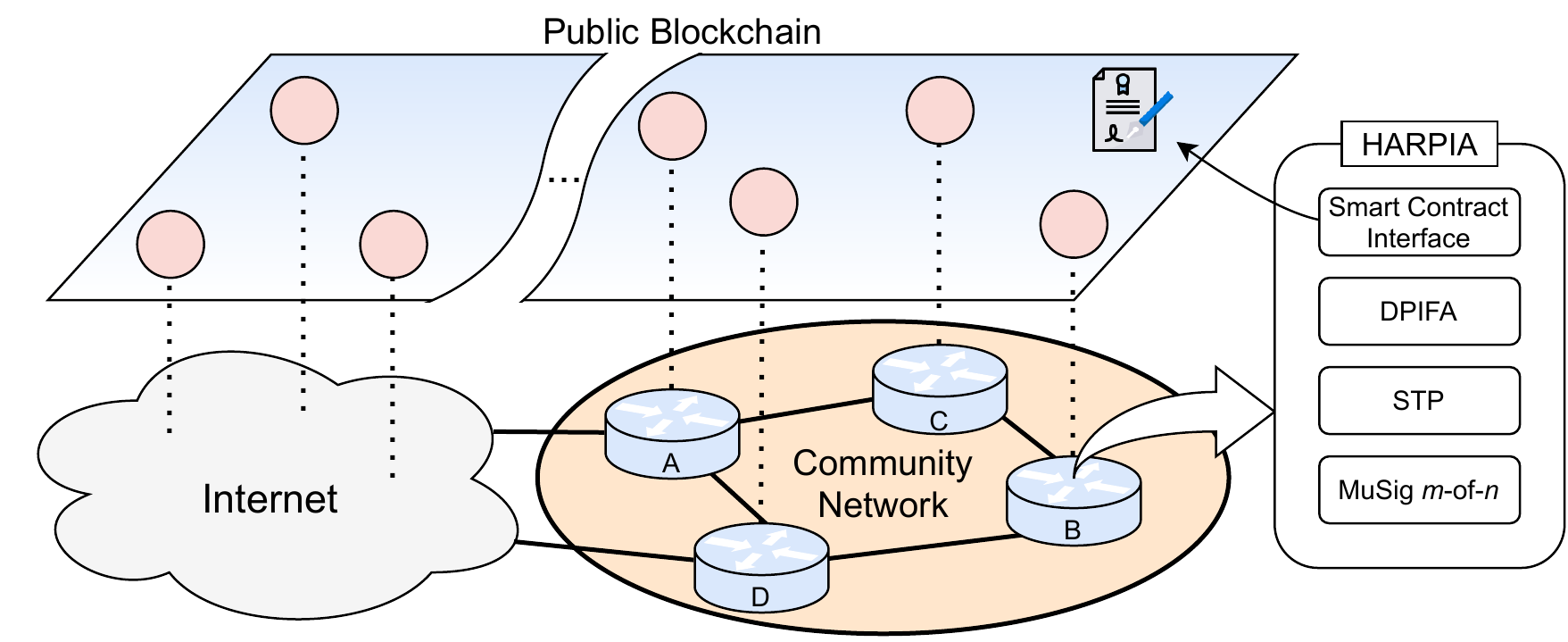}
\caption{HARPIA architecture}
\label{fig:architecture}
\end{figure}

This paper proposes the \emph{Hop-by-hop Accounting and Rewards for Packet
dIspAtching} (HARPIA) (Fig.~\ref{fig:architecture}), a system for
credit-based data forwarding incentive mechanisms built on top of the Ethereum
blockchain. HARPIA relies on network traffic accounting based on a
decentralized version of the PIFA from Yoo \emph{et al.}~\cite{pifa2005}
called DPIFA. Instead of relying on TRSM or TTP, routers
share network accounting reports so that any router can validate its
credibility. HARPIA does not require frequent on-chain transactions, but only
periodic ones (e.g., hourly, daily, weekly, monthly),
secured by $m$-of-$n$ MuSig, that settles pending debts correspondent to
routers' utilization and contribution in data forwarding.

The goal of this work is to present HARPIA and investigate
its advantages, limitations, and applications. Our contribution is a new
blockchain-enabled system for credit-based incentive mechanisms
suitable for community networks. The main advantages of HARPIA compared to
related works (Tab.~\ref{tab:summary}) are the
independence of a TTP or TRSM and fewer on-chain transactions.

The rest of the paper is organized as follows. The next section summarizes the
fundamental concepts used in this paper. Section~\ref{sec:harpia} presents the
HARPIA architecture and its components. Section~\ref{sec:eval} is a
performance analysis of computing, storage, communication, and cryptocurrency
costs involved in HARPIA. Section~\ref{sec:related} presents a comparative
analysis of related works. Section~\ref{sec:sec} is a security threat
assessment. Finally, Section~\ref{sec:conc} shows our conclusions and future
works.

\section{Preliminaries}
\label{sec:pre}
This section introduces the technologies and
cryptographic building blocks of HARPIA. First, we summarize blockchain
and smart contract concepts. Second, we describe the work from Yoo \emph{et al.}~\cite{pifa2005} that
inspired DPIFA. Finally, we present MuSig and how to build a $m$-of-$n$
multi-signature scheme.

\subsection{Blockchain and smart contracts}
\label{sec:bcsm}
Blockchains are distributed databases organized as sequential chains of blocks
secured using cryptographic techniques. Each
ordered block contains transactions validated by a consensus protocol,
e.g., Proof of Work (PoW), Proof of Stake (PoS), Proof of Authority (PoA), or byzantine algorithms. Each participant can
trace and verify the correctness of the data. Recorded data is hard to tamper
with,
ensuring immutability and irreversibility. Bitcoin was the first blockchain
that proposed a secure and P2P system independent of a trusted third-party for payments over the
Internet~\cite{nakamoto2008}. It secures transactions of the most used and
valuable cryptocurrency (BTC) today and inspired many other systems for
innovations other than financial transactions over the Internet.

Blockchains such as Bitcoin also allow encoding rules and scripts for
processing transactions. This feature evolved to support
programs called smart contracts~\cite{smartcontracts2020}.
The consensus protocol automatically enforces smart contracts' trusted execution
in a traceable and irreversible way. Ethereum~\cite{ethereum2021},
for example, transforms blockchains in finite state machines, which state
transitions are equivalent to cryptocurrency transactions, enabling secure
and decentralized applications.

\subsection{PIFA}
\label{sec:pifa}
The Protocol Independent Fairness Algorithm (PIFA)~\cite{pifa2005} is a
data forwarding incentive mechanism for computer networks. The algorithm assumes that
routers do not know full path information from source to destination of packets
so that it could be applied independently of the routing protocol.

\begin{table}[!htb]
\caption{PIFA report fields}
\label{tab:pifafields}
\begin{tabularx}{\columnwidth}{lp{7.5cm}}
\toprule
        \textbf{Field} &
        \textbf{Description} \\
\midrule
        \textbf{RID} & ID of the reporter router\\
\addlinespace
        \textbf{NID} & ID of the neighbor router\\
\addlinespace
        \textbf{I} & Number of input packets from the neighbor\\
\addlinespace
        \textbf{O} & Number of output packets to the neighbor\\
\addlinespace
        \textbf{S} & Number of packets starting at the current router\\
\addlinespace
        \textbf{T} & Number of packets terminated at the current router\\
\addlinespace
        \textbf{OFN} & Number of packets originated from the neighbor\\
\bottomrule
\end{tabularx}
\end{table}

The original version of PIFA has a central service called Credit Manager (CM)
that receives network traffic reports from all network routers. Periodically,
every reporter router (RID) sends one traffic report for each of its interfaces to
the CM. Each message has the traffic statistics with a specific neighbor
router (NID) within a period identified by a sequence number (SEQ). The report fields are
described in Tab.~\ref{tab:pifafields}.
With the aggregated reports, CM can infer the current network topology
and the credibility of individual reports based on the following three
criteria (considering that
$\textrm{Q}_{n,m}$ denotes the $\textrm{Q}$ field of a report where $n$ = RID and $m$ = NID are
neighbors).

\paragraph{$\textrm{O}_{n,m} = \textrm{I}_{m,n}$} The number of output packets from a router
must be the same as the
number of input packets to the opposite side router for every link.

\paragraph{$\textrm{F}_n$}
\label{par:criteriab}
The number of packets forwarded by a router $n$ with $\textrm{A}_n$
neighbors must be the
same as the number of input packets that have not terminated on it and the
        number of output packets that have not been originated from it:
\begin{equation}
    \sum_{m \in \textrm{A}_n} \textrm{I}_{n,m} - \sum_{m \in \textrm{A}_n}
    \textrm{T}_{n,m} = \sum_{m \in \textrm{A}_n} \textrm{O}_{n,m} - \sum_{m \in \textrm{A}_n} \textrm{S}_{n,m}
\end{equation}

\paragraph{$\textrm{OFN}_{m,n} = \textrm{S}_{n,m}$} Finally, the value
$\textrm{S}$ from a router should be
identical to the $\textrm{OFN}$
of the next-hop router. $\textrm{OFN}$ aims to prevent a malicious router from
manipulating $\textrm{F}_n$ by changing both
$\sum \textrm{T}$ and $\sum \textrm{S}$ in criteria (b). This $\textrm{OFN}$ is computed
by counting the number of packets originated at a neighbor out of all
input packets from that neighbor.

CM increases the
credit for each router in proportion to $\textrm{F}_n$, and all routers have to pay these
credits as much as the number of packets they originate. In networks using
PIFA, routers would not deliberately avoid packet forwarding because they need to earn
credits to send their packets.

\subsection{MuSig}
\label{sec:musig}
MuSig~\cite{musig2019} is a Schnorr-based multi-signature scheme that allows a
group of signers to produce a short and joint signature on a common message.
It is provably secure under the discrete logarithm assumption and in the
plain public-key model (meaning that signers are only required to have a
public key, but do not have to prove knowledge of the private key
corresponding to their public key to some certification authority or other
signers before engaging the protocol). The scheme also allows public key
aggregation meaning that verifiers do not need all public keys used in the
signature. These properties are helpful to provide compact multi-signatures in blockchain features
such as multi-party transactions and permissioned consensus.

MuSig is parameterized by group parameters $(\mathbb{G},p,g)$
where $p$ is a k-bit integer, $\mathbb{G}$ is a cyclic group of order $p$,
and $g$ is a generator of $\mathbb{G}$, and by hash functions
($\textrm{H}_{com}$, $\textrm{H}_{agg}$, $\textrm{H}_{sig}$, $\textrm{H}_{tree}$) from $\lbrace 0,1 \rbrace ^{*}$ to
$\lbrace 0,1 \rbrace^{\mathscr{l}}$. The multi-signature process is divided
into three rounds as follows:

\paragraph{Round 1} A group of $n$ signers wants to cosign a message $m$. Let $X_1$ and $x_1$ be
the public and private keys of a
specific signer, let $X_2, ..., X_n$ be the public keys of other cosigners, and let $\langle L \rangle$ be the
multi-set of all public keys involved in the signing process. For $i\in \lbrace 1,...,n \rbrace$, the signer computes
$a_{i} = \textrm{H}_{agg}(\langle L \rangle,X_{i})$ as well as the aggregated public key

\begin{equation}
\label{eq:aggpubkey}
\tilde{X} = \prod_{i=1}^{n}X_{i}^{a_{i}}.
\end{equation}

\paragraph{Round 2} The signer generates a random private nonce $r_{1}\leftarrow\mathbb{Z_{p}}$,
computes $R_{1} = g^{r_{1}}$ (the public nonce), the commitment $t_{1} =
\textrm{H}_{com}(R_{1})$, and sends $t_{1}$ to all other cosigners. When
receiving the commitments $t_{2},...,t_{n}$ from all other cosigners, the
signer sends $R_{1}$ to them. This procedure ensures that the public
nonce is not exposed until all commitments have been received. Upon receiving
$R_{2},...,R_{n}$ from other cosigners, the signer verifies that
$t_{i}=\textrm{H}_{com}(R_{i})$ for all $i\in  \lbrace 2,...,n \rbrace$. The
protocol is aborted if this is not the case.

\paragraph{Round 3} If all random public nonces can be verified using the commitments, then compute
\begin{equation}
    R = \prod^{n}_{i=1}R_{i},
\end{equation}

\begin{equation}
    c = \textrm{H}_{sig} (\tilde{X},R,m),
\end{equation}

and

\begin{equation}
    s_{1} = r_{1} + c a_{1} x_{1} \mod p.
\end{equation}

The signature $s_{1}$ is sent to all other cosigners.
When receiving $s_{2},...s_{n}$ from other cosigners, the signer can compute

\begin{equation}
    s = \sum_{i=1}^{n}s_{i} \mod p.
\end{equation}

Then, the signature is

\begin{equation}
\sigma = (R,s).
\end{equation}

\paragraph{Verification} Given a signed message $m$ and the respective
aggregated public key $\tilde{X}$ (or the set of public keys to
compute $\tilde{X}$ with Eq.~\ref{eq:aggpubkey}), in order
to verify if a multi-signature $\sigma$ is valid, a verifier can compute

\begin{equation}
c =  \textrm{H}_{sig} (\tilde{X},R,m),
\end{equation}

and accepts $\sigma$ if

\begin{equation}
    g^{s} = R\tilde{X}^{c}.
\end{equation}

\paragraph{$m$-of-$n$} MuSig also allows implementing a threshold policy where $m$ valid signatures
of $n$ possible ones are required to confirm a multi-signature.
This feature can be implemented by building a Merkle hash tree~\cite{merkle2004} where the leaves are
permitted combinations of public keys (in the aggregated form), and the nodes
are H$_{tree}$ hashes. The Merkle tree should
be a complete binary tree. Thus the last aggregated public key combination is repeated until the
number of leaves is a power of 2.
The verification, in this case, would take as input an aggregated public key $\tilde{X}$, a
multi-signature $\sigma = (R,s)$ and a Merkle proof $P$. Its validity
would depend on the multi-signature
being valid with the provided key, and the proof establishing that the key is,
in fact, one of the leaves of the Merkle tree, identified by its root hash $\mathcal{M}_{root}$.
This approach is only possible when the number of combinations of aggregated
public keys is feasible, given a $m$-of-$n$ threshold

\section{HARPIA}
\label{sec:harpia}
The \emph{Hop-by-hop Accounting and Rewards for Packet dIspAtching} is a
system that implements a
data forwarding incentive mechanism suitable for community networks built on top of
DPIFA, $m$-of-$n$ MuSig, and Ethereum smart contracts.
The next subsections present an overview of the mechanism and detail each component of HARPIA.

\subsection{Overview}
\label{sec:overview}
In our proposal, routers that originate packets should pay
the next-hop routers that forward their packets. Next-hop routers would not
deliberately avoid data forwarding because their owners want to be paid. HARPIA does
not distribute tokens to routers. Routers should earn cryptocurrency tokens
beforehand through Ethereum mining, an advertising model for content
downloaded, raising donations, or any other means. HARPIA is
focused on the dealings among infrastructure routers and do not impose any limitations for the number of client devices inside
the local network of each infrastructure router.

DPIFA is the component responsible for the traffic accounting. It is inspired
in PIFA but does not require a trusted third-party
(CM). Instead, all routers share accounting reports and calculate how much cryptocurrency each router would
pay or receive.
Like PIFA, HARPIA is routing protocol-agnostic and does not require
topology information to be known. It does not require that all the network
routers participate in the system. However, only HARPIA members can engage in traffic
accounting and cryptocurrency dealings. Additionally, HARPIA members must be
network neighbors to produce credible traffic accounting.

\begin{figure}[!htb]
\centering
    \includegraphics[width=\columnwidth]{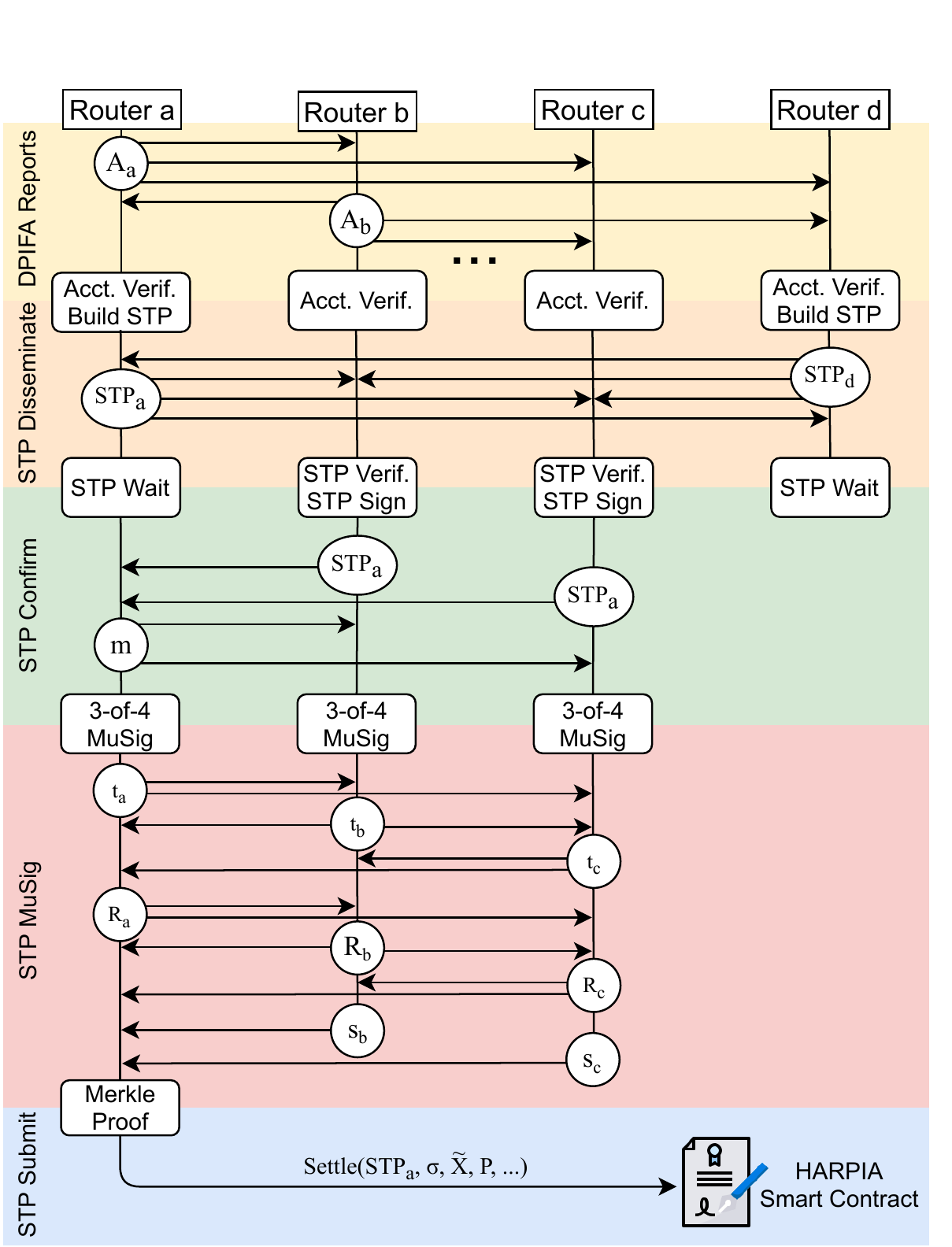}
\caption{Messages in a HARPIA cycle}
\label{fig:pifamusig}
\end{figure}

HARPIA performs an on-chain settlement of pending debts on each HARPIA cycle.
Each cycle takes at least $\beta$ Ethereum blocks mined to complete considering an
average block time $\gamma$.
Each router aggregates DPIFA reports within a cycle to validate their
credibility and calculate pending debts. Thus, any router can propose a
transaction to update the cryptocurrency balance of all routers. If at least
$m$-of-$n$ routers sign this transaction, the proposer can call a smart
contract function that verifies its validity and settles pending debts.
Fig.~\ref{fig:pifamusig} illustrates the messages in a typical
cycle in a network with 4 routers. These messages are described in
the next subsections, except for MuSig, already explained in
Section~\ref{sec:musig}.

\subsection{DPIFA accounting reports}
\label{sec:dpifa}
Each router keeps track of network traffic statistics for sent and
received unicast packets according to DPIFA parameters. Periodically, e.g.,
every $\lambda =$ 10 minutes, each router disseminates DPIFA accounting
reports to all other routers and keep the accounting
information since the last HARPIA settlement.
Each message corresponds to statistics on
a neighbor link and has the same fields as the original PIFA shown in
Tab.~\ref{tab:pifafields}.

After receiving DPIFA reports from the other routers, each router can
calculate how much data each other router forwarded and how much data
originated from it have been forwarded by neighbor routers. If a router loses
reports at the current HARPIA cycle due to communication errors, router
failure, or even malicious packet discarding, it can request the reports
directly from their originators.

\subsection{HARPIA STP}
In HARPIA, payments are not performed directly between pairs of neighbor
routers. HARPIA computes DPIFA aggregated reports to
calculate the total of cryptocurrency tokens that each router $n$ would pay or
receive ($\textrm{C}_{n}$) according to their consumption and contribution in data
forwarding. Any router could create a Settlement Transaction Proposal (STP)
and disseminate it on every HARPIA cycle. In
Fig.~\ref{fig:pifamusig}, routers $a$ and $d$ create and disseminate an STP.
An STP includes a $\textrm{C}_n$ entry for each router $n$ calculated as shown in
Eq.~\ref{eq:pifapricesum} where: $\textrm{A}_n$ is the set of adjacent routers of router $n$;
$\textrm{F}_{n,m} = \textrm{I}_{n,m} - \textrm{T}_{n,m}$ is the amount of data received by
router $n$ (NID) from router $m$ (RID) that has been forwarded toward the
destination;
$\textrm{S}_{n,m}$ represents the data produced by router $n$ and
sent to router $m$ through a link $n$-$m$; $\textrm{H}_{avg}$ is the average hop count between two
routers; and $\textrm{P}_{avg}$ is the average link price. For each link
$n$-$m$ in the network, both routers $n$ and $m$ should specify the same link price
($\textrm{P}_{n,m}$) in the smart contract, so that any router can
calculate the STP and all the routers can validate it.

\begin{equation}
\label{eq:pifapricesum}
    \textrm{C}_n = \sum_{a \in \textrm{A}_n} \textrm{F}_{n,a} \cdot \textrm{P}_{n,a} - \textrm{S}_{n,a} \cdot \textrm{P}_{avg} \cdot \textrm{H}_{avg}
\end{equation}

The STP also includes a reward for its creator with a twofold
goal. It serves as an incentive for the creator to perform the operation and
compensate for the public blockchain transaction costs spent in mining.

All routers can verify the fairness of an STP using the traffic accounting
credibility criteria described in Section~\ref{sec:pifa} and the price of the
network links $P_{n,m}$ of each router. Every router that agrees with an STP
sends back a confirmation using a signed acknowledge message to the STP
creator. In Fig.~\ref{fig:pifamusig}, routers $b$ and $c$ confirm STP$_a$, and router $a$
answers with a message $m$ that informs which routers confirmed the STP.
A minimum percentage of members ($\zeta$) should confirm an STP to engage in
MuSig. In Fig.~\ref{fig:pifamusig}, $\zeta = 75\%$, equivalent to a $3$-of-$4$
routers. If MuSig succeeds, the STP creator can send the transaction to the
Settle smart contract function described in the next subsection. HARPIA MuSig
uses the elliptic curve secp256k1 and the hash algorithm SHA-256.

HARPIA admits an error in accounting up to a
percentage threshold $\delta$. For each router $p$ that creates and
disseminates an STP$_p$, every other router $l$ checks whether every
$\textrm{C}_{n}^{p}$ from this STP is within a local calculated $\textrm{C}_{n}^{l} \pm \delta$.
Furthermore, our scheme allows each router to validate STPs according to other arbitrary
criteria, such as router reputation mechanisms.

\subsection{HARPIA smart contract}
\label{sec:smartcontract}
The central component of our system is a smart contract built on top of
the Ethereum blockchain using the Solidity language v0.7.5. A HARPIA smart contract
instance for a network (from now on, we call it just as a HARPIA instance)
manages membership and cryptocurrency dealings. The
identification of each router member is its Ethereum
account public address. Tab.~\ref{tab:harpiaparameters}
shows a list of HARPIA parameters set in the smart contract.

\begin{table}[!htb]
\caption{HARPIA parameters stored in the smart contract}
\label{tab:harpiaparameters}
\begin{tabularx}{\columnwidth}{lp{6.5cm}}
\toprule
        Parameter &
        Description \\
\midrule
        $\beta$ & HARPIA cycle, in number of Ethereum blocks\\
\addlinespace
        $\lambda$ & DPIFA report period\\
\addlinespace
        $\zeta$ & $m$-of-$n$ MuSig percentage threshold\\
\addlinespace
        $\textrm{P}_{n,m}$ & Price (tokens/GB) for traffic from router $m$ to $n$\\
\addlinespace
        $\delta$ & Tolerated error for each $\textrm{C}_n$ in an STP\\
\addlinespace
        $\xi$ & Duration of an STP, in number of Ethereum blocks\\
\addlinespace
        $\tau$ & Minimum ether balance to keep as a member\\
\addlinespace
        $\phi$ & Minimum ether deposit to Join an instance\\
\addlinespace
        $\mathcal{M}_{root}$ & Merkle tree root of current $\tilde{X}$ combinations\\
\bottomrule
\end{tabularx}
\end{table}

\paragraph{Initialization and membership}
To enable HARPIA in a network, a router must be an Ethereum full node and deploy a HARPIA smart
contract instance on the blockchain. The deployment includes an implicit Join operation that
assigns this router as the first member of this HARPIA instance. After that,
the smart contract allows any neighbor router in the network to join this HARPIA
instance using the Join function. Join and Leave operations are administrative
tasks that require interaction among different routers' owners for approval.
This interaction is a reasonable requirement for community networks because physical link
establishment among the routers also depends on owners' negotiation. Join
operations also require an initial deposit of ether, which minimum value
$\phi$ is defined in
the smart contract deployment. This deposit is converted into HARPIA tokens
valid within the smart contract instance.

\paragraph{Settle}
A Settle function call validates and executes an STP. It can be called once per
HARPIA cycle. At the end of the Settle function, any pending Join or Leave operation
is completed by updating the Merkle tree root of current $\tilde{X}$ combinations ($\mathcal{M}_{root}$) in the smart contract.
Pending Leave operations also withdraw the ether respective to the remaining
HARPIA tokens for the router's Ethereum account. HARPIA limits the
number of Join and Leave pending operations to one per Settle call. This is an
acceptable limitation for community networks because they are relatively
static when compared to other \emph{ad hoc} networks.

An STP is valid for a time corresponding to $\xi$ Ethereum blocks. Thus, the
number of Ethereum blocks mined since the STP has been proposed and the Settle
operation could not exceed $\xi$, otherwise, the Settle call fails.

Initially, each ether unit in the smart contract instance corresponds to one
token, but the Settle call creates new tokens to reward the STP proposer, changing the rate between ether and
tokens. This procedure works also as an inflation mechanism and inhibit
inactive members in the HARPIA instance because idle tokens devalue every
settlement, and could trigger an automatic Leave in the Settle call when a router's ether balance
reaches $\tau$. Thus, HARPIA tends to keep only members that actively use the
network, provide data forwarding services or create HARPIA STPs for the Settle
function.

\paragraph{MuSig verification}
Join, Leave and Settle function calls require a MuSig verification. This
requirement means
these transactions should precede a 3-round MuSig involving at least
$m$-of-$n$ of the current HARPIA instance members. The smart contract
deployment defines the $m$-of-$n$ threshold $\zeta$ for MuSig. $\zeta$ is a
number in the interval $[50-100)$ representing the minimum percentage of
member signatures required in a MuSig. A threshold $\zeta$ is satisfied by any
$m$-of-$n$ scheme that $\frac{100 m}{n} \ge \zeta$ (e.g., if $\zeta$ =
75\%, the schemes $3$-of-$4$, $4$-of-$5$, and $5$-of-$6$ satisfy $\zeta$). The
call for functions that require a MuSig verification should include the
multi-signature $\sigma$ for the transaction parameters, the
respective aggregated public key $\tilde{X}$, and the Merkle proof $P$. The
call validates MuSig in two steps. First, it validates $\tilde{X}$ using $P$
and $\mathcal{M}_{root}$. Second, it
validates $\sigma$ using $\tilde{X}$. The implicit Join in the smart contract
deployment does not require this verification, and Joins in HARPIA instances with
only one member require only a simple signature of the current member instead of a
MuSig. Furthermore, every time a router performs a Join or a Leave,
all valid combinations of aggregated public keys and respective Merkle tree
should be calculated using current members' public keys. This is a computation
intensive operation that should be executed by all members individually.
The router calling Join/Leave is responsible for writing a new $\mathcal{M}_{root}$
in the smart contract.

\paragraph{Other functions} The smart contract also provides functions that do not require MuSig and allow: 1) to read information
about current router members in a HARPIA instance; 2) to
register/unregister/update router links
and respective prices; 3) to buy/redeem tokens for a router so that it can pay for data
forwarding.

\section{Evaluation}
\label{sec:eval}
This section presents an evaluation of computational and cryptocurrency costs
involved in HARPIA. We evaluated processing, storage and communication
performance using Python, Gnuplot and R to identify potential bottlenecks.
Also, we calculated Ethereum smart contract's gas requirements to estimate
cryptocurrency costs. Through this evaluation we discuss the applicability of
the system.

\begin{figure}[!htb]
\centering
\includegraphics[width=\columnwidth]{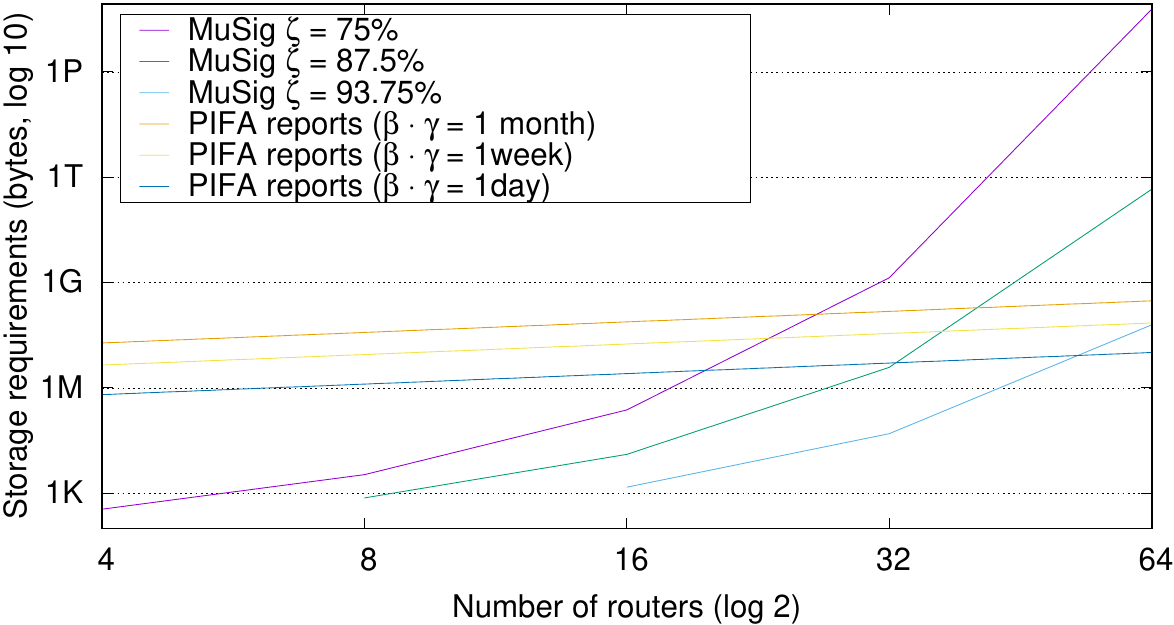}
\caption{Storage requirements for MuSig aggregated public keys with respective
    Merkle hash tree and DPIFA report messages}
\label{fig:storage}
\end{figure}

\begin{figure}[!htb]
\centering
    \includegraphics[width=\columnwidth]{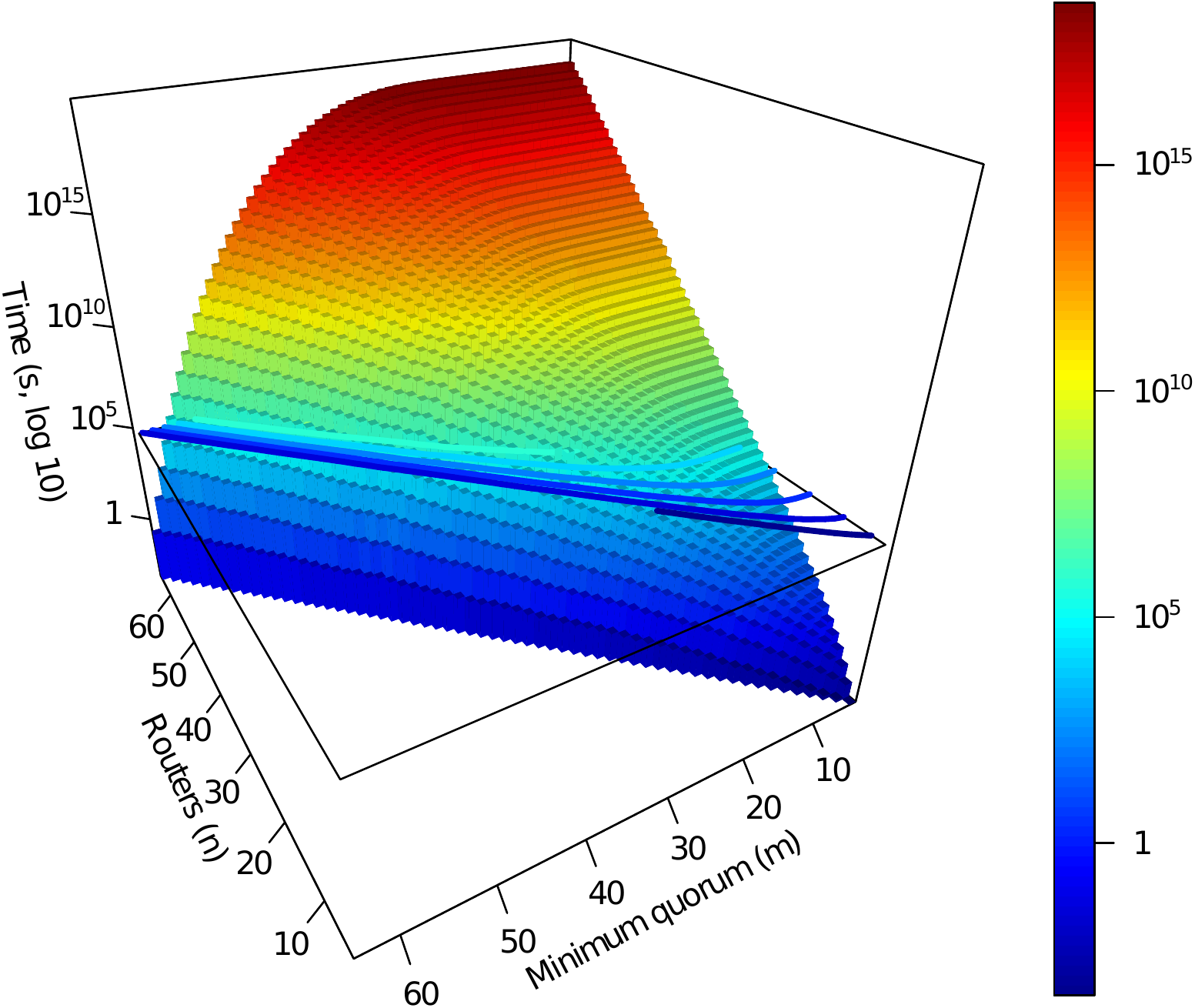}
    \caption{Estimated times to calculate $\mathcal{C}_{\tilde{X}}$ and Merkle hash tree}
\label{fig:rplots}
\end{figure}

The main bottleneck of HARPIA resides in the creation of the aggregated public
keys of all valid combinations of public
keys ($\mathcal{C}_{\tilde{X}}$)~\footnote{$\mathcal{C}_{\tilde{X}} = \sum_{k=m}^{n}\frac{n!}{k! (n-k)!}$
possible ${\tilde{X}}$ in MuSig $m$-of-$n$}
and all $\textrm{H}_{tree}$ (SHA-256) in the Merkle hash tree~\footnote{$2
\mathcal{C}_{\tilde{X}}-1$ $\textrm{H}_{tree}$ operations in the Merkle hash
tree} required for Join and Leave. We evaluated the execution times using off-the-shelf hardware~\footnote{Single threaded Python
script. Linux Kernel 5.4.0, glibc 2.31,
Python 3.8.5 w/ fastecdsa 2.1.5. CPU Intel Core i7-8550U 4GHz, Cache L2 8MB, RAM
16GB, SSD storage w/ 31.6 Gb/s speed.}. Tab.~\ref{tab:evalcpu} shows results
using chosen thresholds $\zeta$ and Fig.~\ref{fig:rplots} shows
estimated times based on partial executions. Results show feasible execution times for specific
thresholds. For example, if we set HARPIA cycle time ($\beta \cdot \gamma$) to
1 day, then any threshold $\zeta$ with execution time under 1 day would be
acceptable.
A plane with level curves is fixed at 16 hours ($\approx 10^{4.76}$ seconds) execution time in
Fig.~\ref{fig:rplots}.
If we arbitrarily set 16 hours as a limit, any valid $m$ and $n$ that result
in execution times under the plane are acceptable $m$-of-$n$ configurations.

\begin{table}[!htb]
    \caption{Execution times to calculate $\mathcal{C}_{\tilde{X}}$ and respective Merkle hash tree}
\label{tab:evalcpu}
    \begin{tabularx}{\columnwidth}{p{2.2cm}XXXX}
\toprule
        \multirow{2}{*}{$\zeta$ = min($m$-of-$n$)} & \multicolumn{4}{r}{Avg. of 30 execs. for each number of routers}\\
     \cmidrule{2-5}
     & 8       & 16     & 32       & 64 \\
\midrule
        $75\%$    & 0.308s   & 39.17s & -- & -- \\
\addlinespace
        $87.5\%$  & 0.100s   & 2.490s  & 1547s & -- \\
\addlinespace
        $93.75\%$ & -- & 0.335s  & 21.15s & 53458s \\
\bottomrule
\end{tabularx}
\end{table}

Fig.~\ref{fig:storage} illustrates how the storage costs scale in HARPIA with
the number of routers. MuSig storage
requirements\footnote{\label{footmusig}$33 \mathcal{C}_{\tilde{X}} + 32 (2
\mathcal{C}_{\tilde{X}} - 1)$. 33 bytes for each $\tilde{X}$ (elliptic curve
point in the compact form) and 32 bytes for each $\textrm{H}_{tree}$ (SHA-256).} are
shown for different $m$-of-$n$ thresholds $\zeta$. DPIFA storage
requirements\footnote{\label{footdpifa}$116 \nu \cdot n \frac{\gamma \cdot
\beta}{\lambda}$. Each DPIFA report message with 116 bytes: 44 bytes for DPIFA
fields, 4 bytes for a timestamp, 4 bytes for a
nonce, and 64 bytes for the digital signature.}
are shown
for different cycle
times $\beta \cdot \gamma$
considering an average of $\nu=5$ neighbors and period $\lambda=10$ minutes. The evaluation shows that HARPIA is suitable for
community networks up to 64 infrastructure routers with attached storages to
keep MuSig and DPIFA data. In a realistic scenario ($n = 32$, $\beta \cdot
\gamma = 1$ week, and $\zeta = 87.5\%$), HARPIA requires 36MB for
DPIFA records\textsuperscript[\ref{footdpifa}], 3.83MB for
MuSig\textsuperscript[\ref{footmusig}], and
419GB\textsuperscript[\ref{foot2}] for the Ethereum
blockchain.

A typical HARPIA cycle disseminates around 5.24MB of messages
(Fig.~\ref{fig:pifamusig}) from
DPIFA, STP and MuSig among all routers considering a community network with 64
routers, cycle time $\beta \cdot \gamma$ = 1 day, period $\lambda$ = 10
minutes, and an average of $\nu$ = 5 neighbors. For bigger networks,
processing times and storage space requirements become a bottleneck before the network
overhead.

\begin{figure}[!htb]
\centering
    \includegraphics[width=\columnwidth]{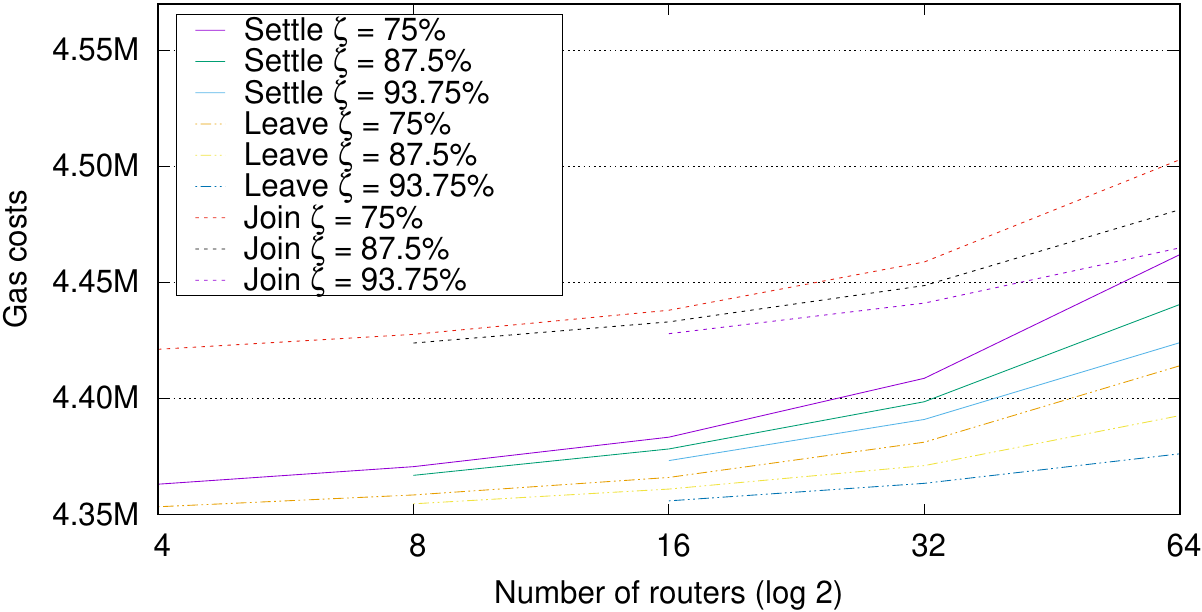}
    \caption{HARPIA gas costs}
\label{fig:gas_settle}
\end{figure}

\begin{table}[!htb]
    \caption{HARPIA gas costs in US\$\textsuperscript{a} (April, 28 2021)}
\label{tab:ethereumcosts}
\begin{threeparttable}
    \begin{tabularx}{\columnwidth}{p{3.4cm}rr}
\toprule
        Gas cost & Cost in Mainnet & Cost in Classic\\
\midrule
        $\approx$ 4.52M~\tnote{d} & 269.41 & 0.73\\
\bottomrule
\end{tabularx}
\begin{tablenotes}
\item[a]{Cost in US\$: Gas cost $\times$ Gas price\textsuperscript{b} $\times$ Exchange rate\textsuperscript{c}}
\item[b]{Gas price: $\approx$ 22 Gwei\textsuperscript{e} in Mainnet and $\approx$ 4.8 Gwei\textsuperscript{e} in Classic}
\item[c]{Exchange rate: 1 ETH = US\$ 2,709.34 and 1 ETC = US\$ 34.02}
\item[d]{Gas limit: $\approx$ 14.9M in Mainnet and $\approx$ 8M in Classic}
\item[e]{1 Gwei = 1 ether $\times$ $10^{-9}$}
\end{tablenotes}
\end{threeparttable}
\end{table}

Smart contracts' execution consumes ether, the Ethereum cryptocurrency. The
amount of ether that a function requires is measured in gas units. Gas
prices oscilate according to blockchain transaction's demand. It is crucial to estimate
gas costs involved in the smart contract because the
Ethereum blockchain imposes a gas limit for functions and to figure out
the scenarios that HARPIA is applicable.
We used Remix~\footnote{https://remix.ethereum.org/} to calculate smart
contract gas costs, and obtained gas prices, gas limits and exchange rates from
Tokenview\footnote{\label{foot2}https://tokenview.com}.
Fig.~\ref{fig:gas_settle} shows gas
costs for functions that require MuSig for different $\zeta$ values. We can
identify that the number of routers in a HARPIA instance has little
influence on gas costs. Tab.~\ref{tab:ethereumcosts} shows gas costs
for function calls converted to US\$ using
current exchange rates for Ethereum Classic (ETC)
and Ethereum Mainnet (ETH). Estimations indicate that the HARPIA smart
contract has relatively high gas costs in the Mainnet but low costs
in Classic. HARPIA applicability would depend on the community network budget
and the frequency of smart contract calls. This frequency can be adjusted with a longer HARPIA cycle
that reduces the number of Settle, Join and Leave operations.

\section{Security threats}
\label{sec:sec}
The main goal of this paper was the presentation of HARPIA and an
evaluation of its applicability. However, there is a series of security
threats specific to our protocol that we discuss here.

\paragraph{Key management} HARPIA does not need to rely on PKI or any other
trusted third-party. Instead, the router's public keys (for ECDSA and MuSig)
must be registered in the smart contract on Join operations. Join
operations require a MuSig involving current HARPIA instance members. This
MuSig depends on an admission process (for example, during a periodic meeting
of current community network members) that validates new members' public keys.
Details of such an admission process are outside the scope of this paper.
Also, MuSig is provably secure under the plain public-key model, and then members
do not have to prove knowledge of the private key corresponding to their
public key.

\paragraph{Malicious DPIFA reports} Similarly to PIFA, we assume that truth-telling will be
the dominant strategy because otherwise, neighbor routers will
not engage in the HARPIA instance. Routers will leave the system if they do
not have their data properly forwarded or their data forwarding services
correctly paid. Any router can verify accounting correctness using aggregated
DPIFA reports. We also assume that there is no collusion among routers.
Nevertheless, we digitally sign (ECDSA) DPIFA records with the Ethereum private key of its
creator using timestamps and nonces to prevent replay attacks.

\paragraph{Traffic spoofing} A router can spoof the source address from its
packets to cheat the next-hop router to believe that those packets
have originated from another router. The goal is to avoid these
packets from being counted as their traffic on DPIFA reports, thus, avoiding
paying the next-hop router. These inconsistencies will influence in DPIFA report
validations. Nonetheless, an additional countermeasure can be applied for this specific
threat. Routers could implement traffic authentication using a public/private
key scheme. In this case,
routers should register the public keys of their interfaces on the HARPIA smart
contract so that all the other routers can authenticate their traffic.

\paragraph{Malicious STPs} We assume that a majority of honest routers will
never validate unfair STPs. It is not in the interest of
routers to produce malicious STPs because they want their STPs to be credible
and validated by other routers to receive a reward for their
proposals. For the same reason, we assume that routers will not engage in
producing valid STPs and maliciously aborting the process at any stage after
the 3-round MuSig begins. A malicious router would adhere to this behavior
to undermine a HARPIA instance postponing Settle operations
indefinitely. However, it is more likely that the router engages in
submitting the transaction on-chain as soon as possible to be rewarded.
Nevertheless, STPs contain the current Ethereum block hash and are digitally
signed using timestamps and nonces to prevent replay attacks.

\paragraph{Quorum for MuSig} There is a risk that the $m$-of-$n$ MuSig quorum
cannot be achieved because part of the routers or respective network links are
faulty or even because they lost their private keys. To prevent this
situation, the inflation mechanism in the Settle operation removes from
the HARPIA instance routers which cryptocurrency balance is under a
pre-defined threshold $\tau$.

\section{Related works}
\label{sec:related}
This section highlights key aspects of representative works in
blockchain-enabled data forwarding incentives for computer
networks~\cite{machado2021} summarized in
Tab.~\ref{tab:summary}. Those works have been designed for community
networks~\cite{meshdapp2019,althea2020}, device-to-device (D2D)
networks~\cite{lot492019}, delay-tolerant networks (DTN)~\cite{truthful2018} and the
Internet~\cite{routebazaar2015,trautmann2020}.

\begin{table*}
\small
\centering
\caption{Related works in blockchain-enabled data forwarding incentives for computer networks}
\label{tab:summary}
\begin{tabularx}{\textwidth}{llllll}
\toprule
System & Blockchain & Network & Payment proof & Forwarding proof & TTP\\
\midrule
Truthful Inc.~\cite{truthful2018} & Bitcoin & DTN & On-chain PoW & Receipts & No \\
\addlinespace
MeshDapp~\cite{meshdapp2019} & Ethereum & Com. Net. & On-chain PoA & Traf. acct. & Oracle service \\
\addlinespace
RouteBazaar~\cite{routebazaar2015} & Bitcoin & Internet & On-chain PoW & GRE tun. acct. & Intermediate ASs\\
\addlinespace
Althea~\cite{althea2020} & Cosmos & Com. Net. & On-chain PoS & VPN tun. acct. & Peers and exit nodes \\
\addlinespace
LOT49~\cite{lot492019} & Bitcoin & D2D & Off-chain Lightning & Receipts & Witness nodes \\
\addlinespace
Rout. Based Blockc.~\cite{trautmann2020} & -- & Internet & On-chain PoR & PoR & No \\
\midrule
HARPIA & Ethereum & Com. Net. & On-chain PoW & DPIFA & No\\
\bottomrule
\end{tabularx}
\end{table*}

Most of the works deal with
the free-riding problem similarly to pre-blockchain credit-based incentive
mechanisms. In this context, packet forwarding is a service rewarded with
cryptocurrency. We name as \emph{payment proofs} and \emph{forwarding proofs} the methods for assuring that a party paid and the
other party performed the correspondent packet forwarding correctly.
However, we understand that some
mechanisms unify these proofs, i.e., the same mechanism provides
both proofs.

Payment proofs are secure blockchain transactions for cryptocurrency transfers that fall into two
categories:
on-chain~\cite{routebazaar2015,truthful2018,meshdapp2019,althea2020} or
off-chain~\cite{lot492019}. The first consists in typical blockchain
transactions that require mining (e.g., PoW) or
validation (e.g., PoS or PoA) for consensus. Usually, validation allows faster (more transactions per
second) and cheaper (fewer cryptocurrency fees) transactions.
Off-chain transactions consists in blockchain scalability
solutions~\cite{zhou2020} such as channels and childchains.

Forwarding proofs can be classified according to two criteria. First,
according to the mechanism itself: traffic
monitoring~\cite{routebazaar2015,meshdapp2019,althea2020} or cryptographic receipts in the
network protocol~\cite{truthful2018,lot492019}. Second, whether
they need a trusted third-party~\cite{routebazaar2015,meshdapp2019,althea2020,lot492019} or
not~\cite{truthful2018,trautmann2020}.

RouteBazaar~\cite{routebazaar2015}, MeshDapp~\cite{meshdapp2019} and
Althea~\cite{althea2020} need to trust in the agents that perform traffic
accounting. In RouteBazaar, the intermediate autonomous systems (AS) monitor
traffic and perform expensive on-chain accounting. MeshDapp relies on an
oracle service responsible for the network traffic monitoring. In Althea,
every pair of neighbors should perform traffic monitoring for its neighbors'
network interfaces. Also, Althea has a method to detect fraud in neighbors'
accounting that uses tunnels to exit nodes on the Internet.

Receipt-based forwarding proofs can be combined with on-chain~\cite{truthful2018} or
off-chain transactions~\cite{lot492019}. While LOT49~\cite{lot492019} supports
cheap off-chain transactions with the Lightning protocol micropayments, it
requires a witness node that acts as a trusted third-party. Trautmann and
Burnell’s patent~\cite{trautmann2020} describes a system that introduces a
Proof of Routing (PoR) scheme that can securely implement a blockchain network and
provide useful consensus. Their blockchain-based router idea includes
different nodes that process data packets between endpoints to produce
cryptographic proofs, similarly to PoW schemes. Nodes can be router nodes,
which analyze and route data packets, or block nodes that manage collections
of specially labeled packets and generate new blocks in the blockchain.

\section{Conclusions and future works}
\label{sec:conc}
This paper presented HARPIA, a blockchain-enabled system for
credit-based incentive mechanisms for data forwarding in computer networks. Unlike
related works, HARPIA does not require frequent on-chain transactions, and is
independent of a TTP or TRSM. HARPIA is built on top of DPIFA
traffic accounting
and Ethereum smart contract transactions secured by $m$-of-$n$ MuSig.

Our evaluation shows that lower $m$-of-$n$ MuSig thresholds increase
processing, network, and storage requirements with combinatorial complexity.
Also, DPIFA storage requirements increase linearly with shorter periods and
longer cycles. The analysis reveals that HARPIA is suitable for community
networks with up to 64 infrastructure routers with thresholds above 75\% and
periods greater than 10 minutes. These limitations are acceptable since a
considerable number of community networks have less than 64 routers (e.g.,
60\% of Freifunk list~\footnote{https://freifunk.net/}). HARPIA smart contract's gas
costs are relatively high in the Ethereum Mainnet and present affordable fees
in the Ethereum Classic. Gas costs can also be reduced increasing the cycle to
produce fewer MuSig signed transactions.

For future works, we will simulate representative community networks'
topologies under different HARPIA parameters emulating selfish and
malicious routers. We also plan to investigate other threshold multi-signature
schemes, such as those presented by Boneh \emph{et al.}~\cite{boneh2018} and
Nick \emph{et al.}~\cite{musig2}, to evaluate the most appropriate alternatives.

\end{document}